\documentclass{WileyMSP-template}
\usepackage{amsmath}
\usepackage{amsfonts}
\usepackage{amssymb}
\usepackage{xcolor}

\def\beq{\begin{equation}}
\def\eeq{\end{equation}}
\def\beqa{\begin{eqnarray}}
\def\eeqa{\end{eqnarray}}

\begin{document}

\pagestyle{fancy}
\rhead{\includegraphics[width=2.5cm]{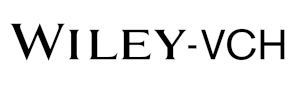}}

\title{Updated bounds on Axion-Like Particle Dark Matter with the optical MUSE-Faint survey}

\maketitle


\author{Elisa Todarello}


\begin{affiliations}
Dipartimento di Fisica, Universit\`a di Torino\\
via Giuria 1,  I–10125 Torino, Italy\\
Istituto Nazionale di Fisica Nucleare, Sezione di Torino\\ 
via P. Giuria 1, I–10125 Torino, Italy\\
Email Address: elisamaria.todarello@unito.it\\

\end{affiliations}



\begin{abstract}

Bounds are derived on the axion-like particle (ALP) to two-photon coupling in the mass range $2.65-5.27$~eV.
The bounds are obtained by searching for the signal from ALP decay in the Multi Unit Spectroscopic Explorer (MUSE) observations of five dwarf spheroidal galaxies, under the assumption that ALPs constitute the dark matter component of the haloes.
These bounds are of the same order and improve on the robustness of those of Reference~\cite{Regis}, and currently represent the strongest bounds within the considered mass range.
\end{abstract}


\section{Introduction}
Axion-like particles (ALPs), just like QCD axions, couple to photons through the interaction Lagrangian
\beq
\mathcal{L} = -\frac{1}{4} g a F_{\mu\nu}\tilde{F}^{\mu\nu} \enspace,\label{lagrangian}
\eeq
where $\tilde{F}^{\mu\nu}$ is the dual of the electromagnetic field strength tensor $F^{\mu\nu}$, $a$ is the ALP (or axion) field, and $g$ is a coupling constant with dimensions of inverse energy. The interaction described by Equation~\eqref{lagrangian} is used in a wide range of laboratory, cosmological, and astrophysical searches for ALPs, see References~\cite{Irastorza:2018dyq, DiLuzio:2020wdo, Sikivie:2020zpn} for reviews. One process that can occur as a result of this interaction is the decay of an ALP into two photons, which we use in this work to establish upper bounds on $g$.

We search for the diffuse photon emission from ALP decay in five dwarf galaxies targeted by the MUSE-Faint survey~\cite{museI}: Eridanus 2, Grus 1, Hydra II,  Leo T, and Sculptor. Dwarf satellite galaxies are ideal targets for dark matter searches, thanks to their high light-to-mass ratio and proximity to Earth. 
We selected these particular five dwarfs because information on the dark matter profile, needed to calculate the expected flux, is available either from the MUSE collaboration itself~\cite{museIII} or can be found in the literature.

The photons from ALP dark matter decay have an energy of about half the ALP mass. Since the MUSE instrument measures in the optical band \cite{musedeep}, we can constrain $g$ for masses between 2.65 and 5.27~eV. Other bounds in a similar mass range have been derived from stellar evolution in globular clusters~\cite{GC1, GC2}, other telescope searches for ALP decay focused on galaxy clusters~\cite{VIMOS}, and anisotropies in the cosmic optical background~\cite{COB1, COB2}.

In this work, we update the bounds presented in Reference~\cite{Regis}, which were obtained from the MUSE observation of only one dwarf, namely Leo T. Using multiple targets enables us to establish more reliable limits. All five dwarfs provide comparable bounds of the order of $g < 10^{-12}$~GeV$^{-1}$, increasing our confidence in the validity of such a constraint. We also provide combined bounds, which are slightly more stringent than those from any of the single dwarfs.

The analysis presented here does not take into account the uncertainty on the dark matter profile, whose parameters we fix to the best-fit values form~\cite{museIII} or, in the case of Sculptor, from a new analysis to be presented in an upcoming publication~\cite{us}. While Ref.~\cite{us} will include the likelihood of the dark matter profile in the statistical analysis, the present manuscript offers a more pedagogical discussion of the bounds and their derivation.

\section{Observations and data from the MUSE instrument}

The Multi Unit Spectroscopic Explorer (MUSE) located at the Very Large Telescope (VLT) is a large-field medium-resolution Integral Field Spectrograph. It provides data in the optical band in the wavelength range $470 - 935$~nm, with a sampling of 1.25~\AA\, for a total of 3721 channels. The spectral resolution has a full width at half maximum $\Delta\lambda$ varying with wavelength between $2.5$ and $3.0$~\AA. The data is in the form of a ``cube", the first two dimensions being the angular coordinates on the sky and the third dimension being the wavelength.
The field of view is $1\times 1$~arcmin$^2$ with a spatial resolution of $\sim0.6$~arcsec.

Known emission lines from the night sky are subtracted from the data using the MUSE Data Reduction Software (DRS)~\cite{museI}. The subtraction cannot be executed perfectly, something we take  into account in our analysis by adding an arbitrary spatially flat term to the expected flux, as emission from the night sky is not expected to vary over the angular scales of the image.
The DRS also provides the r.m.s. error estimate that we'll use for our data analysis.

For each galaxy, we mask in every channel the brightest sources, such as stars. To create the mask we run SExtractor~\cite{SExtractor} on the white-light image, i.e. the image obtained by integrating the cube over the wavelength axis.

\section{Expected flux}
The rate of ALP decay into two photons is 
\begin{equation}
\Gamma = \frac{g^2m^3}{64\pi}\enspace,
\end{equation}
where $m$ is the ALP mass. Assuming spherical symmetry, the expected flux density  at the detector from a direction $\theta$ on the sky is 
\begin{equation}
S_\lambda (\theta)=\frac{\Gamma}{4\pi}\,\frac{1}{\sqrt{2\pi}\sigma_\lambda} e^{-\frac{(\lambda-\lambda_{obs})^2}{2\sigma_\lambda}}\int d\Omega\, d\ell\, \rho[r(\theta, \Omega, \ell)]\,B(\Omega) \enspace,\label{flux}
\end{equation}
where $\rho(r)$ is the dark matter energy density as a function of the distance from the center of the dwarf, $\lambda_{obs}$ is twice the Compton wavelength for the ALP mass under consideration, Doppler shifted as described below,  $\sigma_\lambda = \Delta\lambda / (2\sqrt{2\ln(2)})$ is the spectral resolution, and $B(\Omega)$ is the detector's angular response. The integral is taken over the line of sight $\ell$ and the position inside the angular beam $\Omega$.

In the ALP rest frame, the photons are emitted back to back each carrying energy $m/2$.
In the lab frame, the emission line is smeared out due to the dark matter velocity dispersion in the dwarf, and Doppler-shifted because of the relative velocity between the dwarf and Earth. 
For the galaxies we consider, the velocity dispersion is ${\delta v\lesssim 10^{-4}}$, while the resolution in wavelength of the MUSE instrument is $\sigma_\lambda/\lambda \geq 1.2\times 10^{-4}$. We thus neglect the velocity dispersion and assume that all decay photons have a frequency of $m/2$ when emitted. 

We cannot instead neglect the radial velocities of the dwarfs. These velocities are reported in Table~\ref{tab1} of the Appendix. When converted to natural units, we see that for all dwarfs except Leo T, the radial velocity is well above the best spectral resolution. Therefore,  photons from ALP decay would be detected in a different frequency channel than the one with frequency $m/2$. To account for this, in our calculation of the bounds presented below, we apply a correction for each dwarf
\beq
\nu_{em} = \nu_{obs} (1 + v_{radial}) \enspace,
\eeq
where $\nu_{obs}$ is the frequency of the MUSE channel being considered, and $\nu_{em}$ is the corresponding frequency at emission. We then report the bounds as a function of the ALP mass $m = 4\pi\nu_{em}$.

\section{Data analysis and results}

We assume a cored dark matter density profile for the five dwarfs. The parameters are taken to be the best-fit values from the GravSphere  Jeans analysis of~\cite{museIII} for Eridanus 2, Grus 1, Hydra II, and Leo T, while for Sculptor we use the best-fit values from an analysis to be presented in~\cite{us}. This new analysis makes use of data from~\cite{Walker, Coleman_2005}. The expression for the energy density as a function of radius, as well as the complete set of parameter values, are presented in the Appendix.

For our statistical analysis, we use a methodology completely analogous to that of~\cite{Regis}.
For each galaxy, and for each frequency channel, we compute the likelihood 
\beq 
\mathcal{L}=e^{-\chi^2/2} \;\;\; {\rm with} \;\;\; \chi^2=\frac{1}{N_{pix}^{FWHM}}\sum_{i=1}^{N_{pix}} \left(\frac{S_{th}^i-S_{obs}^i}{\sigma_{rms}^i}\right)^2\enspace.\label{chi2}
\eeq
The sum runs over pixels in the image. It is divided by the number of pixels contained within the detector's angular resolution $N_{pix}^{FWHM}$, as the fluxes in these pixels are correlated.
We assume that the theoretically predicted flux density $S_{th}^i$ at a wavelength $\lambda$ consists of a flat component $S_{\lambda, flat}$ constant across the image plus $S_\lambda^i$ from Equation~\eqref{flux}. $S_{\lambda, flat}$ accounts for imperfect subtraction of emission from the night sky. 
The noise estimate $\sigma_{rms}^i$ is derived through the MUSE DRS. For each channel, we scan a grid of values of $\Gamma$ and $S_{\lambda, flat}$.

\begin{figure}
\centering
  \includegraphics[width=0.7\linewidth]{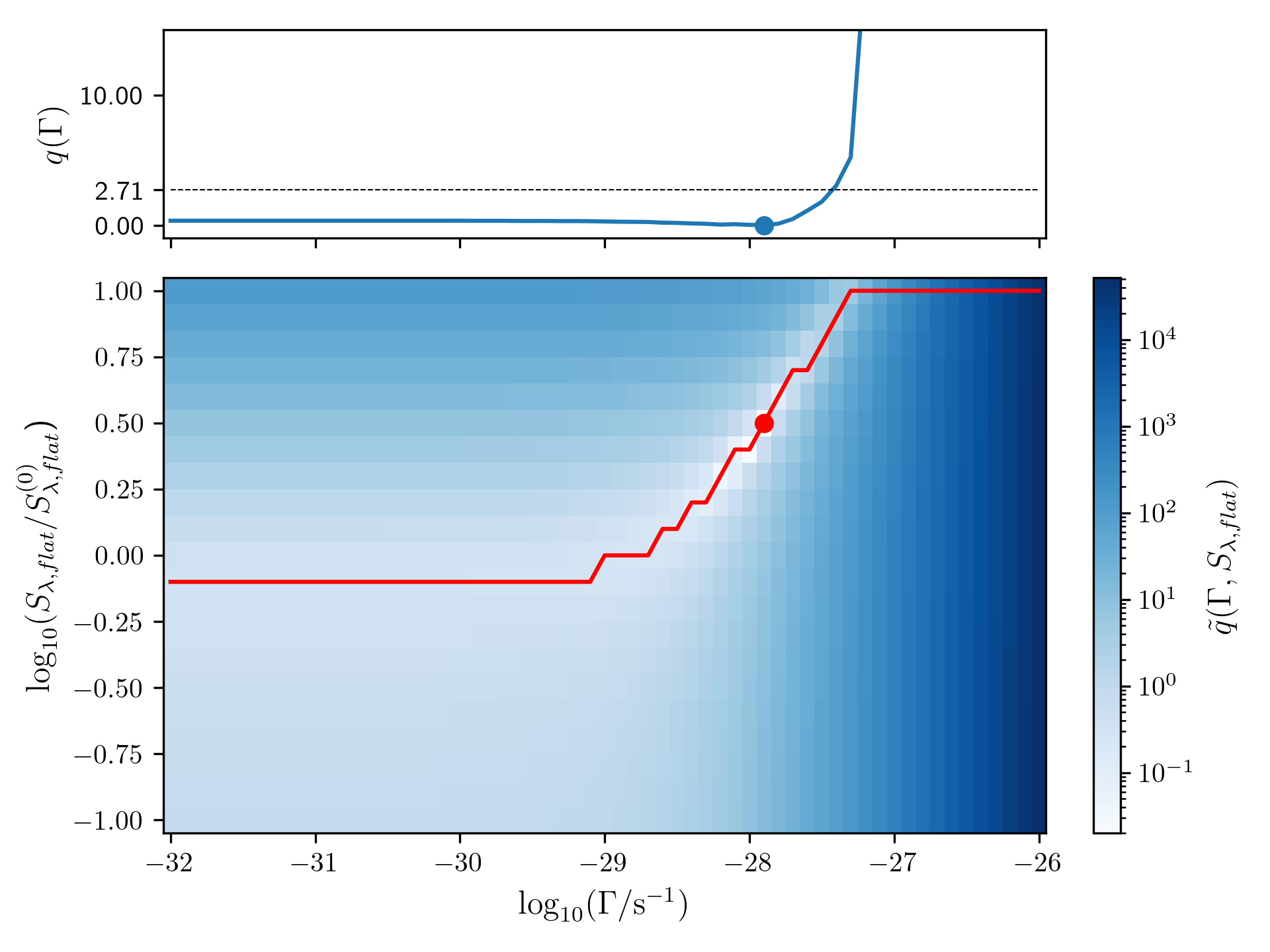}
  \caption{Illustration of the profile likelihood methodology used to obtain upper bounds on $g$, shown for the channel of wavelength 5325~\AA~of Leo T as an example. On the bottom panel,  the color gradient represents the value of $\tilde{q}(\Gamma, S_{\lambda,flat})$. On the vertical axis, $S^{(0)}_{\lambda,flat}$ is the best-fit value of $S_{\lambda,flat}$ in the absence of dark matter. We use $S^{(0)}_{\lambda,flat}$ as an initial guess to determine the range of values to scan.
  The red line marks, for each $\Gamma$, the lowest value of $\tilde{q}$, while the red dot marks the best fit $\hat{\Gamma},~\hat{S}_{\lambda, flat}$. 
  Notice that, for this channel, $S^{(0)}_{\lambda,flat}$ is negative. As we increase $\Gamma$, larger negative values of $S_{\lambda,flat}$ are preferred to compensate for the fact that the diffuse emission from dark matter is not a good fit.
  The top panel shows the resulting profile likelihood $q(\Gamma)$ (solid blue). Again, the dot  marks the best-fit value $q(\Gamma)=0$, while the intersection between the thin dotted line and the blue line corresponds to the 95\% C.L. upper bound on $\Gamma$ and thus on $g$.}
  \label{fig:profile}
\end{figure}

\begin{figure}[t]
\centering
  \includegraphics[width=0.8\linewidth]{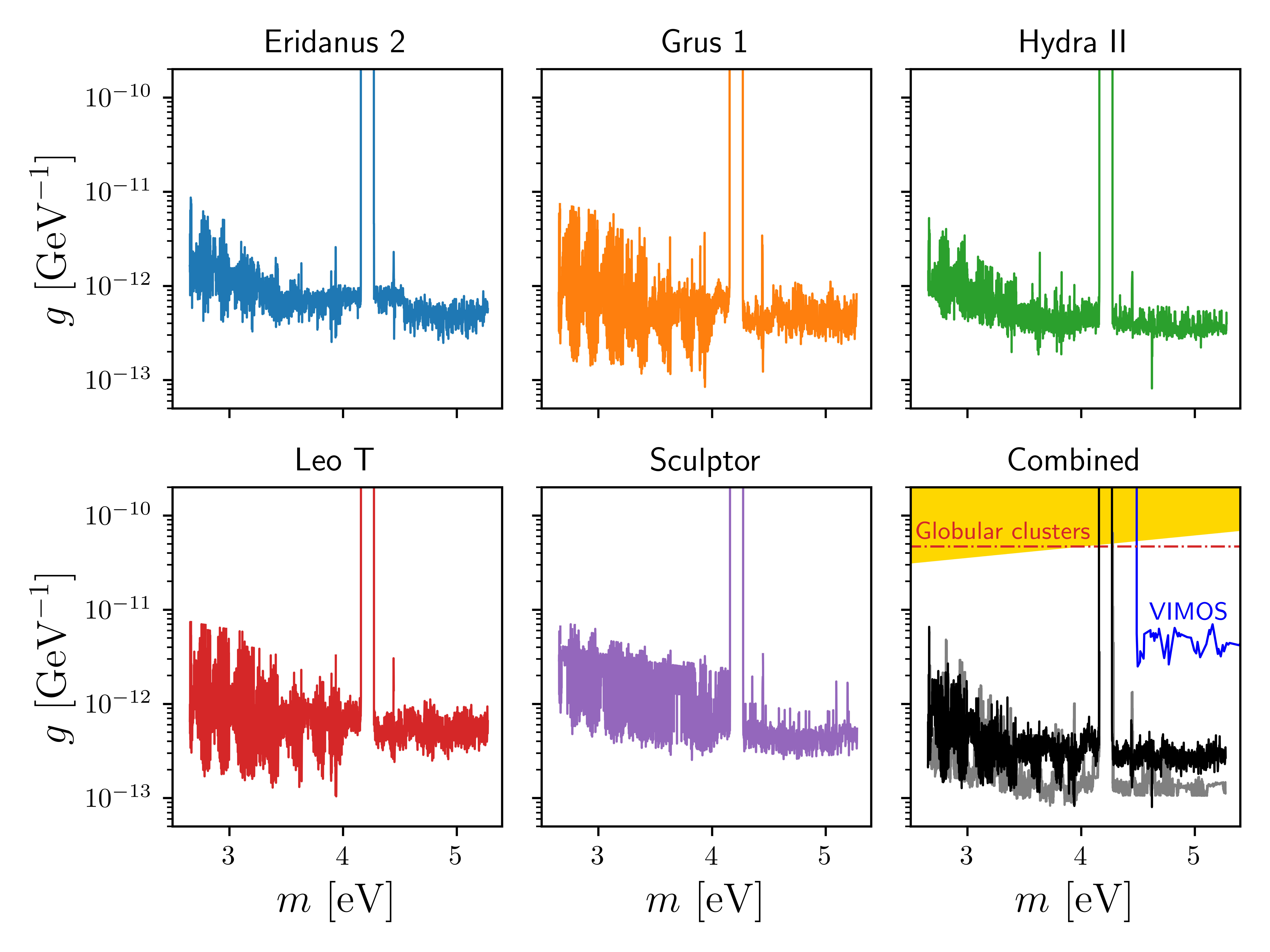}
  \caption{Upper bounds on the ALP-photon coupling $g$ at the 95\% C.L. derived from MUSE-Faint observations of five dwarf spheroidal galaxies assuming a `coreNFWtides' profile, and combined bounds (solid black, bottom-right panel). On the combined panel we show the QCD axion band~\cite{DiLuzio:2016sbl} in yellow. We also show bounds from star evolution in globular clusters from~\cite{GC2} (red dash-dotted line) and from observation of galaxy clusters~\cite{VIMOS} (blue solid line). The grey line on the combined panel represents the bounds of~\cite{Regis} derived assuming an NFW profile for Leo T and using the best-fit D-factor from~\cite{Bonnivard} (see main text for details).}
  \label{fig:bounds}
\end{figure}
To establish our bounds, we choose a frequentist approach that does not require incorporating prior knowledge on the allowed parameter ranges. We are concerned with constraining the ALP-photon coupling $g$, while we are not interested in $S_{\lambda, flat}$. A commonly employed technique to deal with such nuisance parameters within the frequentist approach is the profile likelihood. The profile likelihood function is obtained by maximizing the likelihood  with respect to the nuisance parameter for each fixed value of the parameter of interest. This yields a one-dimensional slice through the likelihood surface, as illustrated on the bottom panel of \textbf{Figure~\ref{fig:profile}} for the channel of wavelength 5325~\AA of Leo T. On the axes of the bottom panel, we see the grid of scanned values. The color gradient represents the value of \\
${\tilde{q}(\Gamma, S_{\lambda,flat}) = -2\ln\left(\mathcal{L}(g, S_{\lambda, flat})/\mathcal{L}(\hat{g}, \hat{S}_{\lambda, flat}) \right)}$, where a hat indicates the best-fit value. For each value of $\Gamma$, we select the lowest value of $\tilde{q}(\Gamma, S_{\lambda,flat})$ corresponding to any of the values of $S_{\lambda, flat}$. Such values are marked by the red line. 
The result of the profiling procedure is shown on the top panel of \textbf{Figure~\ref{fig:profile}}. The blue line represents the likelihood ratio
\begin{equation}
q(g) =    -2 \ln\left(\frac{\mathcal{L}(g, \hat{\hat{S}}_{\lambda, flat})}{\mathcal{L}(\hat{g}, \hat{S}_{\lambda, flat})}\right)\enspace,
\end{equation}
where double hat indicates that the parameter has been profiled out. 

We use the following test statistics
\begin{equation}
t(g) =
\begin{cases}
   q(g)&\qquad g\geq\hat{g}\\
  0 &\qquad g<\hat{g}\enspace.\label{teststat}
\end{cases} 
\end{equation}
To obtain upper bounds on $g$, we assume that the probability density function for $t(g)$ is a half $\chi^2$ with one degree of freedom, meaning that the probability of $t(g)$ being larger than a given value $t_c$ is $P(t_c) = \int_{t_c}^\infty  dt' e^{-t'/2} / (2 \sqrt{2 \pi t' }) $. To set the 95\% C.L. upper bound, we search for the lowest value of $g$ for which $t(g)>2.71$. This is also depicted on the top panel of \textbf{Figure~\ref{fig:profile}}.
For the combined bound, we multiply the likelihoods for the single galaxies, here labeled by $j$, yielding the combined likelihood
\begin{equation}
 \mathcal{L}^{tot}(g)=\prod_{j=1}^5 \mathcal{L}^j(g,\hat{\hat{S}}_{\lambda, flat})\enspace,
\end{equation}
where $S_{\lambda, flat}$ has been profiled out for each dwarf independently.

Upper bounds on $g$ at the 95\% C.L. are shown in 
\textbf{Figure~\ref{fig:bounds}}. All five dwarfs give similar bounds in the $10^{-12}$~GeV$^{-1}$ ballpark. The large oscillations for low masses are due to bright atmospheric emission lines that saturate the detector. Since these lines are the same in the observations of all galaxies, they do not compensate and appear in the combined bound as well. Bounds cannot be found for $4.12~\mathrm{eV}<m<4.27~\mathrm{eV}$ because a filter is applied to block light from the sodium laser of the adaptive optic system.

We notice that the bounds from Leo T alone, as well as the combined bounds, are less stringent than those of~\cite{Regis}. This is due to the more conservative dark matter profile assumed here. In Ref.~\cite{Regis}, the profile considered was a Navarro-Frenk-White (NFW), and the integral over the line of sight of the dark matter density of Leo T was derived from the D-factor reported in~\cite{Bonnivard}, while here we're using a `coreNFWtides' with  parameters derived from the MUSE-Faint observations of~\cite{museIII}. These observations reach distances from the center of the  dwarf well below the half-light radius (see for example Figure 2 of ~\cite{museIII}), while the data from~\cite{Bonnivard} are on angular scales of order or larger than the half-light radius. This is one of the factors that makes the present bounds more robust than those of~\cite{Regis}.

\begin{figure}
\centering
  \includegraphics[width=0.8\linewidth]{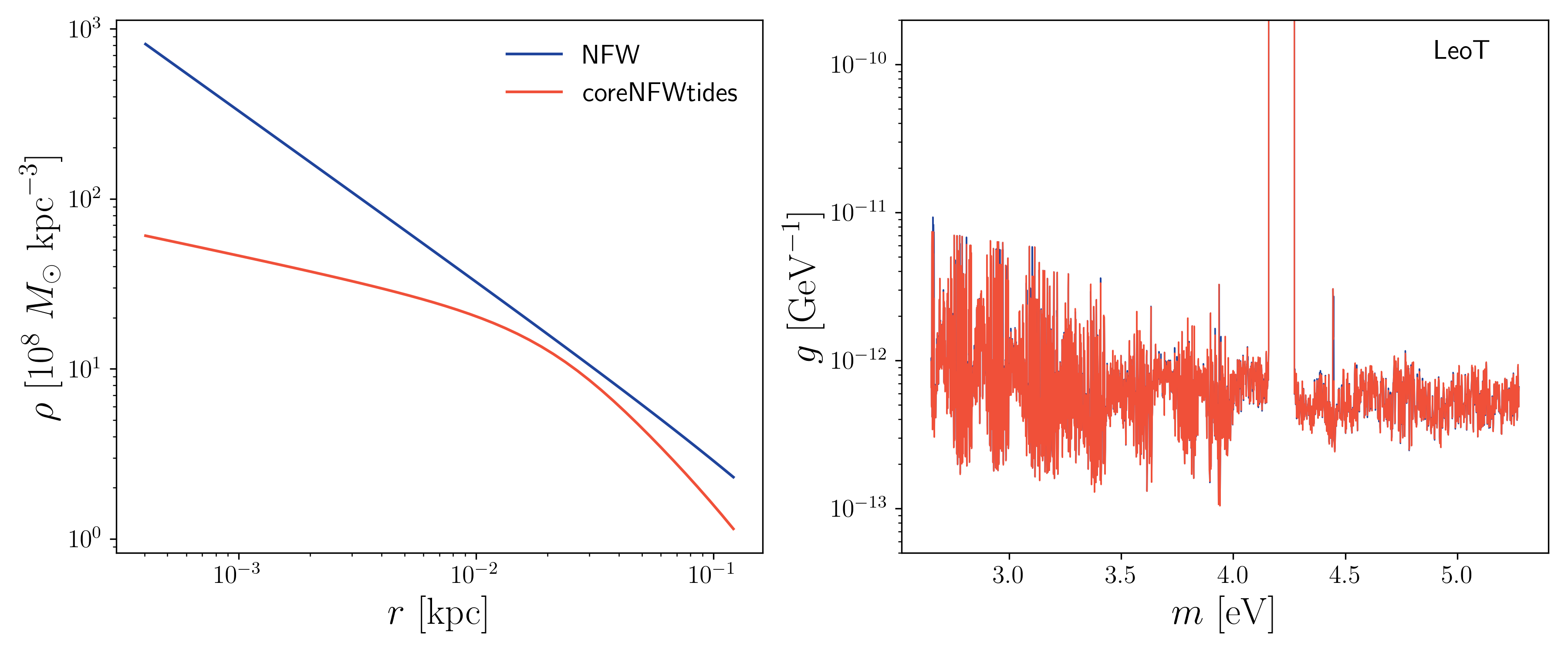}
  \caption{
  Illustration of how our bounds change when the dark matter profile is modified in the central region of the galaxy. On the left panel, we see two possible profiles for the Leo T galaxy, a cuspy NFW and a cored `coreNFWtides'. The maximum radius shown corresponds to our integration limit of 1$^\prime$ from the center of the galaxy. The two profiles are similar for the largest radii. This results in the bounds, shown on the right panel, being almost identical (the blue line is covered by the pink one). See the main text for more details.}
  \label{fig:bounds_cored}
\end{figure}

In order to understand how the bounds change when a cuspy profile is considered, we plot in \textbf{Figure~\ref{fig:bounds_cored}} a comparison between the `coreNFWtides' and NFW profiles for Leo T.
The parameters for the NFW profile are taken from (Table A.1) of~\cite{museIII} and reported in the Appendix for completeness.
On the left panel, we plot both profiles up to our maximum integration radius $r_{max}$, corresponding to an angular size of $1^\prime$. From Equations~\eqref{flux} and~\eqref{chi2}, we see that for fixed $S^i_{obs}$, if we increase $\rho$ in a given pixel, the fit will tend to prefer lower values $\Gamma$, up to variations in $S_{\lambda, flat}$. Next, we notice that the largest the radius, the more pixels associated with it, their number increasing proportionally to $r$. Thus we expect the larger radii to determine the typical variation in the bound on $g$ as $\rho$ is modified. In the example of \textbf{Figure~\ref{fig:bounds_cored}}, the average value of $\rho_{NFW}/\rho_{coreNFWtides}$, weighted by $r$, is approximately 1.7. So we should expect the bounds on $g$ using the cored profile to be worse by a factor of about 1.3. As a matter of fact, the bounds using the two profiles are basically indistinguishable, and we can attribute the bulk of the discrepancy with our estimate to the effect of the nuisance parameter $S_{\lambda, flat}$, which can also compensate for the change in $\rho$. We conclude that if the core is much smaller than $r_{max}$, and if the profiles are similar outside of the core region, the bounds will be roughly unaffected. For larger cores, the bounds will be less stringent for the cored profile by a factor of order $\lesssim \sqrt{\rho_{NFW}(r_{max})/\rho_{core}(r_{max})}$.

Let us finally address the question of whether the data shows any evidence for ALP dark matter. For example, the Leo T channel of \textbf{Figure~\ref{fig:profile}} contains no evidence, as $q$ does not increase significantly for values of $\Gamma$ approaching zero. However, this is not the case for all channels. In fact, for all dwarfs, we find numerous channels in which the null hypothesis is strongly disfavored. This is due, just like the oscillations in the value of the upper bounds, to bright atmospheric emission lines that cannot be perfectly subtracted from the data. This renders the search for evidence of ALPs complicated, as the `fake' evidence due to the atmosphere has to be removed. Such an analysis is in preparation and will appear in~\cite{us}.

\section{Conclusions}
We have provided upper bounds on the axion-like particle to photon coupling by searching for a signal from radiative decay.
Our analysis is based on observations of five dwarf spheroidal galaxies, namely Eridanus 2, Grus 1, Hydra II, Leo T, and Sculptor, obtained from the MUSE-Faint survey.  Unlike Reference~\cite{Regis}, which relied on observations of a single galaxy, we used data from multiple targets, making our constraints more reliable and conclusive.
Additionally, information on the dark matter profile is taken, except for Sculptor, from the very same MUSE-Faint observations~\cite{museIII}, extending to distances significantly shorter than the half-light radius of the dwarf.   This  contributes to the enhanced reliability of the current constraints compared to those presented in the previous work of Ref.~\cite{Regis}.

We have also discussed in detail the method used for our analysis, namely the profile likelihood, and the effect using a cuspy versus a cored profile has on the bounds.

The bounds reported in this manuscript do not consider the uncertainty on the dark matter profile parameters. In a paper currently in preparation, we will take into account the likelihood of the dark matter profile of the dwarfs in the analysis.

\medskip
\textbf{Acknowledgements} \par 
The author would like to thank the members of the team working on this project: Marco Regis, Javier Reynoso, and Marco Taoso from University of Turin and INFN Turin,  Jarle Brinchmann, Daniel Vaz, and Sebastiaan Zoutendijk from the MUSE collaboration. 
The author also thanks the Galileo Galilei Institute for Theoretical Physics for its hospitality during the completion of this work.
The author acknowledges support from the `Departments of Excellence 2018-2022' grant awarded by the Italian Ministry of Education, University and Research (\textsc{miur}) L.\ 232/2016 and Research grant `From Darklight to Dark Matter: understanding the galaxy/matter  connection to measure the Universe' No.\ 20179P3PKJ funded by \textsc{miur}.

\appendix
\section{Dark matter profile parameters}

\begin{table}[h!]
\begin{center}
\begin{tabular}{|| c |  c | c | c | c | c ||} 
 \hline
  & & & & & \\[-2ex]
& Eri 2 & Gru 1 & Hya II & Leo T & Scu \\ [1ex] 
 \hline\hline
 & & & & & \\[-2ex]
$\rho_s~[10^8~M_\odot~\mathrm{kpc}^{-3}]$ & 0.171 & 0.345 & 0.0388 & 0.221 & 0.154 \\ [1ex] 
 \hline
 & & & & & \\[-2ex]
 $r_s~[\mathrm{kpc}]$ & 1.43 &  0.581 &  23.1 &  1.51 &  1.81 \\ [1ex] 
 \hline
  & & & & & \\[-2ex]
 $r_c~[\mathrm{kpc}]$ & 0.0827 &  0.0112 &  0.018 &  0.0217 &  0.0493 \\ [1ex] 
 \hline
  & & & & & \\[-2ex]
 $n~$ & 0.854 &  0.364 &  0.631 &  0.707 &  0.775\\ [1ex] 
 \hline
 & & & & & \\[-2ex]
 $r_t~[\mathrm{kpc}]$ & 1.09 &  3.73 &  16.3 &  9.78 &  1.67 \\ [1ex] 
 \hline
  & & & & & \\[-2ex]
 $\delta$ & 3.35 &  3.79 &  3.94 &  4.20 &  4.22 \\ [1ex] 
 \hline
 & & & & & \\[-2ex]
 $d~[\mathrm{kpc}]$ & 366 & 125 & 151 & 417 & 86 \\ [1ex] 
 \hline
& & & & & \\[-2ex]
 $v_{radial}~[\mathrm{km/s}]$ & 76 & -141 & 303 & 38 & 111 \\ [1ex] 
 \hline
\end{tabular}
\end{center}
\caption{Values of the `coreNFWtides' parameters used for the analysis. They are the best-fit values from the likelihood of the Jeans analysis of~\cite{museIII} using GravSphere
or from an analysis using data from~\cite{Walker, Coleman_2005} in the case of Sculptor.
The table also shows the value of the distance of the dwarfs $d$, taken from~\cite{museIII, museII, McConnachie_2012} and the radial velocities from~\cite{Wenger:2000sw}.
}
\label{tab1}
\end{table}

The bounds shown in \textbf{Figure~\ref{fig:bounds}} are derived assuming the `coreNFWtides' dark matter density profile used for the GravSphere analysis of~\cite{museIII}. 
This profile is constructed starting from an NFW progenitor 
\beq
\rho_{\rm{NFW}}(r)=\frac{\rho_s}{\left(\frac{r}{r_s}\right)\left( 1 + \frac{r}{r_s} \right)^2} \enspace.
\label{eq:rho}
\eeq
and redistributing some of the mass from the central regions to larger radii as follows
\begin{equation}
    \rho_\mathrm{cNFW}(r) = f^n\rho_\mathrm{NFW}(r) + \frac{nf^{n-1} \Bigl(1-f^2\Bigr)}{4\pi r^2r_\mathrm{c}} M_\mathrm{NFW}(<r) \enspace,
\end{equation}
where $M_\mathrm{NFW}(<r)$ is the mass contained within a radius $r$ if the density profile is NFW and $f = \tanh(r/r_\mathrm{c})$.
Moreover, beyond a tidal radius $r_t$, the energy density drops abruptly. The expression for the full profile is
\begin{equation}
\rho_\mathrm{cNFWt}(r) =
\begin{cases}
\rho_\mathrm{cNFW}(r) & r < r_\mathrm{t}\enspace, \\
\rho_\mathrm{cNFW}(r_t)\,(r/r_\mathrm{t})^{-\delta}, & r > r_\mathrm{t} \enspace.
\end{cases}
\end{equation}

The values of the NFW profile parameters of Leo T used to produce \textbf{Figure~\ref{fig:bounds_cored}} are 
\beq
\rho_s = 0.227\times10^8~M_\odot~\mathrm{kpc}^{-3} \qquad r_s =1.45~\mathrm{kpc} \enspace,
\eeq
taken from Table A.1 of~\cite{museIII}.

\medskip

%
\bibliographystyle{MSP}
\bibliography{refs}

\begin{thebibliography}{10}
\providecommand{\url}[1]{\texttt{#1}}
\providecommand{\urlprefix}{URL }

\bibitem{Regis}
M.~Regis, M.~Taoso, D.~Vaz, J.~Brinchmann, S.~L. Zoutendijk, N.~F. Bouch\'e,
  M.~Steinmetz,
\newblock \emph{Phys. Lett. B} \textbf{2021}, \emph{814} 136075.

\bibitem{Irastorza:2018dyq}
I.~G. Irastorza, J.~Redondo,
\newblock \emph{Prog. Part. Nucl. Phys.} \textbf{2018}, \emph{102} 89.

\bibitem{DiLuzio:2020wdo}
L.~Di~Luzio, M.~Giannotti, E.~Nardi, L.~Visinelli,
\newblock \emph{Phys. Rept.} \textbf{2020}, \emph{870} 1.

\bibitem{Sikivie:2020zpn}
P.~Sikivie,
\newblock \emph{Rev. Mod. Phys.} \textbf{2021}, \emph{93}, 1 015004.

\bibitem{museI}
S.~L. {Zoutendijk}, J.~{Brinchmann}, L.~A. {Boogaard}, M.~L.~P. {Gunawardhana},
  T.-O. {Husser}, S.~{Kamann}, A.~F. {Ramos Padilla}, M.~M. {Roth}, R.~{Bacon},
  M.~{den Brok}, S.~{Dreizler}, D.~{Krajnovi\'{}c},
\newblock \emph{A\&A} \textbf{2020}, \emph{635} A107.

\bibitem{museIII}
S.~L. {Zoutendijk}, M.~P. {J{\'u}lio}, J.~{Brinchmann}, J.~I. {Read}, D.~{Vaz},
  L.~A. {Boogaard}, N.~F. {Bouch{\'e}}, D.~{Krajnovi{\'c}}, K.~{Kuijken},
  J.~{Schaye}, M.~{Steinmetz},
\newblock \emph{arXiv e-prints} \textbf{2021}, arXiv:2112.09374.

\bibitem{musedeep}
R.~{Bacon}, S.~{Conseil}, D.~{Mary}, J.~{Brinchmann}, M.~{Shepherd},
  M.~{Akhlaghi}, P.~M. {Weilbacher}, L.~{Piqueras}, L.~{Wisotzki},
  D.~{Lagattuta}, B.~{Epinat}, A.~{Guerou}, H.~{Inami}, S.~{Cantalupo}, J.~B.
  {Courbot}, T.~{Contini}, J.~{Richard}, M.~{Maseda}, R.~{Bouwens},
  N.~{Bouch{\'e}}, W.~{Kollatschny}, J.~{Schaye}, R.~A. {Marino}, R.~{Pello},
  C.~{Herenz}, B.~{Guiderdoni}, M.~{Carollo},
\newblock \emph{A\&A} \textbf{2017}, \emph{608} A1.

\bibitem{GC1}
A.~Ayala, I.~Dom\'\i{}nguez, M.~Giannotti, A.~Mirizzi, O.~Straniero,
\newblock \emph{Phys. Rev. Lett.} \textbf{2014}, \emph{113}, 19 191302.

\bibitem{GC2}
M.~J. Dolan, F.~J. Hiskens, R.~R. Volkas,
\newblock \emph{JCAP} \textbf{2022}, \emph{10} 096.

\bibitem{VIMOS}
D.~Grin, G.~Covone, J.-P. Kneib, M.~Kamionkowski, A.~Blain, E.~Jullo,
\newblock \emph{Phys. Rev. D} \textbf{2007}, \emph{75} 105018.

\bibitem{COB1}
K.~Nakayama, W.~Yin,
\newblock \emph{Phys. Rev. D} \textbf{2022}, \emph{106}, 10 103505.

\bibitem{COB2}
P.~Carenza, G.~Lucente, E.~Vitagliano \textbf{2023}.

\bibitem{us}
E.~{Todarello}, et~al.,
\newblock \emph{in preparation} .

\bibitem{SExtractor}
E.~{Bertin}, S.~{Arnouts},
\newblock \emph{Astron. Astrophys. Suppl.} \textbf{1996}, \emph{117} 393.

\bibitem{Walker}
M.~G. {Walker}, M.~{Mateo}, E.~W. {Olszewski},
\newblock \emph{Astron.J.} \textbf{2009}, \emph{137}, 2 3100.

\bibitem{Coleman_2005}
M.~G. Coleman, G.~S.~D. Costa, J.~Bland-Hawthorn,
\newblock \emph{The Astronomical Journal} \textbf{2005}, \emph{130}, 3 1065.

\bibitem{DiLuzio:2016sbl}
L.~Di~Luzio, F.~Mescia, E.~Nardi,
\newblock \emph{Phys. Rev. Lett.} \textbf{2017}, \emph{118}, 3 031801.

\bibitem{Bonnivard}
V.~Bonnivard, C.~Combet, M.~Daniel, S.~Funk, A.~Geringer-Sameth, J.~A. Hinton,
  D.~Maurin, J.~I. Read, S.~Sarkar, M.~G. Walker, M.~I. Wilkinson,
\newblock \emph{Monthly Notices of the Royal Astronomical Society}
  \textbf{2015}, \emph{453}, 1 849.

\bibitem{museII}
S.~L. Zoutendijk, J.~Brinchmann, N.~F. Bouch\'e, M.~d. Brok, D.~Krajnovi\'c,
  K.~Kuijken, M.~V. Maseda, J.~Schaye,
\newblock \emph{Astron. Astrophys.} \textbf{2021}, \emph{651} A80.

\bibitem{McConnachie_2012}
A.~W. McConnachie,
\newblock \emph{The Astronomical Journal} \textbf{2012}, \emph{144}, 1 4.

\bibitem{Wenger:2000sw}
M.~Wenger, et~al.,
\newblock \emph{Astron. Astrophys. Suppl. Ser.} \textbf{2000}, \emph{143} 9.

\end{thebibliography}

\end{document}